\documentclass[11]{article}
\usepackage{latexsym}
\usepackage{graphics}
\usepackage{epsfig}
\usepackage[dvips]{color}
\usepackage[mathscr]{eucal}
\usepackage{amscd}
\usepackage{amssymb}
\usepackage{amsmath}
\usepackage{amssymb}
\usepackage{amsthm}
\usepackage{geometry}

\newcommand{\pprec}{\prec\!\!\prec}
\newcommand{\ssucc}{\succ\!\!\succ}
\newcommand{\re}{\mathbb R}
\newcommand{\msM}{\mathcal{M}}
\newcommand{\msN}{\mathcal{N}}
\newcommand{\msP}{\mathcal{P}}
\newcommand{\msA}{\mathcal{A}}
\newcommand{\msF}{\mathcal{F}}
\newcommand{\bM}{\mathbb M}

\newcommand{\cT}{\mathcal T}
 
\newcommand{\nin}{\not\in}
\newtheorem{theorem}{Theorem}[]

\newtheorem{lemma}[theorem]{Lemma}
\newtheorem{claim}[theorem]{Claim}
\newtheorem{prop}[theorem]{Proposition}
\newtheorem{corollary}[theorem]{Corollary}

\theoremstyle{definition}
\newtheorem{defn}[theorem]{Definition}

\title{Causal Topology in  \\ Future and Past  Distinguishing
  Spacetimes} \author{Onkar Parrikar\(^{a,b}\) and Sumati
  Surya\(^{c,d}\)\\\\ \(^{a}\){\it \small Birla Institute of Technology and Science
  - Pilani, Goa campus, Goa, 403726, India}  \\  \(^{b}\) {\it  \small 
  Department of Physics, University of Illinois at Urbana-Champaign, 
 Urbana, IL 61801-3080, U.S.A.}\\ \(^{c}\){\it \small Raman Research Institute, CV Raman Ave, Sadashivanagar, 
 Bangalore, 560080, India} \\ \(^{d}\) {\it   \small Department of Physics, McGill University,
 Montr\'eal, QC, H3A 2T8 Canada}}

\begin{document}
\baselineskip=14pt 
\maketitle 

\begin{abstract}
  The causal structure of a strongly causal spacetime is particularly well endowed. Not only does it
  determine the conformal spacetime geometry when the spacetime dimension $n >2$, as shown by
  Malament and Hawking-King-McCarthy (MHKM), but also the manifold dimension. The MHKM result,
  however, applies more generally to spacetimes satisfying the weaker causality condition of future
  and past distinguishability(FPD), and it is an important question whether the causal structure of
  such spacetimes can determine the manifold dimension. In this work we show that the answer to this
  question is in the affirmative. We investigate the properties of future or past distinguishing
  spacetimes and show that their causal structures determine the manifold dimension.  This gives a
  non-trivial generalisation of the MHKM theorem and suggests that there is a causal topology for
  FPD spacetimes which encodes manifold dimension and which is strictly finer than the Alexandrov
  topology.  We show that such a causal topology does exist. We construct it using a convergence
  criterion based on sequences of ``chain-intervals'' which are the causal analogs of null geodesic
  segments.  We show that when the region of strong causality violation satisfies a local
  achronality condition, this topology is equivalent to the manifold topology in an FPD spacetime.
\end{abstract}

\section{Introduction}

In the Riemannian geometry influenced discourse on Lorentzian geometry, the causal structure
$(M,\prec)$ of a spacetime $(M,g)$ is viewed as a derivative construction which relies on the
underlying differentiable structure of $M$, with the causal relation $\prec$ between events on $M$
being obtained from the local lightcone structure provided by $g$.  $(M,\prec)$ is, however, itself
rich with information about the spacetime geometry, and it has been the focus of several
investigations over the years to endow it with a more primitive role in Lorentzian geometry
\cite{KP,gkp,hkm,malament}.  We will concern ourselves in this work only with the causal structure of
{\sl causal} spacetimes, i.e., those that harbour no closed causal curves. For such spacetimes
$(M,\prec)$ is a partially ordered set, i.e., $\prec$ is (i) {\sl acyclic}: for $x,y \in M$, $x\prec
y \prec x \Rightarrow x=y$ and (ii) {\sl transitive}: for $x,y,z \in M$ $x\prec y$ and $y \prec z$
$\Rightarrow x \prec z$. It is important to note that the set of events $M$ in $(M,\prec)$
does not carry with it the attendant differentiable structure of its spacetime avatar.  $(M,\prec)$
has no analog in Riemannian geometry, and it is therefore of very general interest to understand the
role it plays in Lorentzian geometry \cite{KP,gkp,penrose,EH,BE}.  

A set of results due to Malament and Hawking-McCarthy-King \cite{malament,hkm} (MHKM) provides an
important relationship between $(M,\prec)$ and $(M,g)$ in spacetimes that are {\sl future and past
  distinguishing} (FPD). These are spacetimes in which the chronological (time-like) past and future
sets are unique for every spacetime event.  MHKM address the general question: what  aspects of the
spacetime geometry are left invariant under a causal structure preserving bijection between two
spacetimes? Such a map is called a \textsl{causal bijection}.\footnote{The notion of a causal structure preserving map has also been studied from a different perspective in \cite{GPS}, under the name of isocausality class.}

\begin{defn}
A {\sl causal bijection} $f: (M_1,g_1) \rightarrow (M_2,g_2)$ is a bijection
between the set of events $M_1$ and $M_2$ which, in addition, preserves the causal relations $\prec_1$
and $\prec_2$: for $x_1,y_1 \in M_1$, $x_1 \prec_1 y_1 \Rightarrow f(x_1) \prec_2 f(y_1)$ and for
$x_2,y_2 \in M_2$, $x_2\prec_2 y_2\Rightarrow f^{-1}(x_2) \prec_1 f^{-1}(y_2)$.
\end{defn}

Malament's original results were restricted to chronological bijections, i.e., those that preserve only the
chronological relation, but as shown by Levichev \cite{levichev} causal bijections, themselves imply
chronological bijections. We can summarise these results as 

\begin{theorem} {\bf Malament-Hawking-King-McCarthy-Levichev(MHKML)} \\
If a causal bijection $f$ exists between two n-dimensional spacetimes which are both future and
past distinguishing, then these spacetimes are conformally isometric when $n>2$.  \label{MHKMLth}
\end{theorem} 
Here, a {\sl conformal isometry} between $(M_1,g_1)$ and $(M_2,g_2)$ is a bijection
$f:M_1\rightarrow M_2$ such that $f$ and $f^{-1}$ are smooth and $f_*g_1=\Omega^2g_2$ for some real, smooth, non-vanishing function
$\Omega$. Importantly, this implies that $M_1$ and $M_2$ have the same topology.

The power of the MHKML theorem is evident. It tells us that the causal structure $(M,\prec)$ of an
$n$-dimensional spacetime which is future and past distinguishing determines its conformal
geometry and its topology. The {\it only} remaining geometric degree of freedom not determined by the causal
structure is the conformal factor $\Omega$. This suggests an alternative, non-Riemannian approach to
Lorentzian geometry in which a partially ordered set $(M,\prec)$ plays a primitive rather than a
derivative role. Indeed, the MHKML theorem provides a strong motivation for the causal set approach
to quantum gravity in which a locally finite partially ordered set replaces the spacetime continuum
\cite{blms}. 

Can the MHKML theorem be generalised to include spacetimes of different dimensions?  Equivalently,
are causal bijections rigid enough to constrain the spacetime dimension? Let us consider for the moment
a special subclass of FPD spacetimes, namely those that are strongly causal.  For such spacetimes,
the Alexandrov topology $\msA$ is a causal topology which is equivalent to  the manifold topology
$\msM$ (theorem 4.24 in \cite{penrose}), so that $(M,\prec)$ determines the manifold topology and hence
dimension. Since a causal bijection $f$ preserves the chronological relation \cite{levichev} and
hence the topology $\msA$, this implies:
\begin{corollary} \label{cor-one} If there is a causal bijection between two strongly causal
  spacetimes then they both have the same manifold dimension and topology.
\end{corollary}  
This generalises the MHKML theorem when applied to strongly causal spacetimes.  However, since
$\msA$ is known to be strictly coarser than $\msM$ for spacetimes that are not strongly causal, it
is an interesting question whether causal bijections impose topological constraints in such
spacetimes.  In this work, we show that Corollary \ref{cor-one} can be generalised to spacetimes
satisfying weaker causality conditions, i.e., those that are either past {\it or} future
distinguishing:   
\begin{prop} \label{claim-two}\theoremstyle{definition} If there is a causal bijection between two future (or past)
  distinguishing spacetimes then they are of the same dimension.
\end{prop}  
This gives us a genuine generalisation of the MHKML theorem to FPD spacetimes for spacetime
dimension $n>2$. Thus for an FPD spacetime $(M,\prec)$ encodes the manifold topology and hence its
dimension. This begs the question -- what is the causal topology for FPD spacetimes that corresponds
to the manifold topology? We define a new causal topology $\msN$ derived from a
convergence condition on sequences of {\sl chain intervals} which are order theoretic analogs of
null geodesic segments. We show that $\msN$ is equivalent to the manifold topology $\msM$ for an FPD
spacetime in which the regions of strong causality violation satisfy a certain local achronality
condition.

In Section \ref{prelims} we briefly review some of the standard material on causal structure, in the
process formalising many of the concepts and definition given in this section.  We also define and
discuss the properties of {\sl chain intervals} which we use throughout our paper, not only to
construct the new topology $\msN$ but also to give a {\it local} characterisation of strong
causality and future and past distinguishability. In Section \ref{mal} we investigate the properties
of FPD spacetimes in some detail and find some new and generic features. Lemma \ref{claim-two}
follows from this analysis.

In Section \ref{newtop} we define a new convergence criterion on sequences of chain intervals from
which we derive the causal topology $\msN$.  We show that for FPD spacetimes which satisfy an
additional local achronality condition for the regions of strong causality violation, $\msN$ is
equivalent to the manifold topology. This construction is inspired by the work of \cite{malament},
where null geodesic segments are used instead of chain intervals, albeit with an entirely different
motivation.  Indeed, a convergence condition using null geodesic segments instead of chain intervals
always gives the manifold topology for FPD spacetimes, but since null geodesic segments aren't order
theoretically defined, this does not give rise to a purely {\it causal} topology. The construction
of $\msN$ via a convergence criterion however does not give us a causal basis. Finding
such a basis has proved difficult and we end this section with some speculative remarks.

In Section \ref{other} we examine another candidate for a causal topology inspired by a causal
convergence criterion defined in \cite{fullwood}.  The convergence criterion used in \cite{fullwood}
gives rise to a causal topology $\msF$ which is strictly finer than $\msM$ and is Hausdorff and
equivalent to the path topology constructed by Hawking-McCarthy-King \cite{hkm}) iff the spacetime
is past and future distinguishing. $\msF$ can alternatively be defined in terms of its basis
elements, obtained from ``doubled'' chronological intervals and this makes it attractive to work
with. We show that a natural extension of the convergence condition of \cite{fullwood} gives rise to
yet another distinct causal topology $\msP$ which is strictly coarser than $\msF$ and also strictly
finer than $\msM$ when the spacetime is FPD. It is thus tempting to conclude that the the
topological content of $(M,\prec)$ is far richer than one might have imagined and that further work
could yield valuable insights.

\section{Preliminaries} \label{prelims}

We give below the basic definitions and tools  we will need for our paper, and refer the reader to
the classic texts on causal structure \cite{EH,penrose,wald} for a more detailed exposition. 
A spacetime $(M,g)$, is specified by a smooth, $n$-dimensional Hausdorff manifold $M$ with
topology $\msM$ and a smooth Lorentzian geometry $g$. Every $x \in M$ lies in a neighbourhood $N\in \msM$
which is {\sl convex and normal}, i.e., where the exponential map $\exp_p$ is a diffeomorphism from an open
neighbourhood of the origin in Minkowski spacetime $\bM^n$ to $N$ for every $p\in N$. Thus, locally,
the lightcone structure of $(M,g)$ is identical to that of $\bM^n$. Every $x \in M$ also lies in a
{\sl simple region}, i.e., a convex normal open set whose closure lies in a convex normal neighbourhood.

A causal (chronological) curve is a smooth (with respect to $\msM$) map $\gamma: I\rightarrow M$
where $I \subset \re$, such that the tangent to $\gamma$ is everywhere non-spacelike (timelike) with
respect to the metric $g$. A causal or chronological curve is future or past directed depending on
whether its tangent is everywhere future or past directed.  $x$ is said to causally precede $y$ ($x
\prec y$) if there is a future directed causal curve from $x$ to $y$.  Similarly, $x$ is said to
chronologically precede $y$ ($x\pprec y$) if there is a future directed timelike curve from $x$ to
$y$. The {\sl horismotic relation} $x \rightarrow y$ is then defined as $ x \prec y, \, x \not\pprec
y$. We will call a causal curve for which every pair $x \prec y$ is horismotic, a horismotic curve. We
will say that two points $x,y$ are {\sl spacelike related}, or {\sl incomparable}, if $x \not\prec
y$ and $y \not\prec x$.

In the standard usage $\prec$ is  reflexive, i.e. $x \prec x$ and $\pprec$ is irreflexive $x
\not\pprec x$.  The causal future and pasts of an event $x$ are the sets $J^+(x) \equiv \{ y \in M|
y \succ x \} $ and $J^-(x) \equiv \{ y \in M| y\prec x\} $ respectively, and the chronological
future and past are the sets $I^+(x) \equiv \{ y \in M| y \ssucc x \} $ and $I^-(x) \equiv \{ y \in
M| y\pprec x\} $. Both $I^\pm(x) $ are open in the manifold topology. Following \cite{KP} we write
$<x,y>=I(x,y) \equiv \{ z \in M| x \pprec z \pprec y \} $ and $[x,y]=J(x,y) \equiv \{ z \in M | x
\prec z \prec y \} $. The {\sl Alexandrov topology} $\msA$ is generated by the manifold open sets $<x,y>$.

$(M,g)$ is said to be {\sl causal} if $\prec$ is acylic, i.e., $x \prec y$ and $ y \prec x$ implies that
$x=y$. The causal structure $(M,\prec)$ obtained from such a spacetime is then a (reflexive)
partially ordered set, namely, $\prec$ is (i) acyclic and (ii) transitive, i.e., $x \prec y$ and $y
\prec z$ $\Rightarrow x \prec z$. Importantly, $M$ is taken here to be simply the set of events,
without the additional topological and differentiable structures that one needs in defining the
spacetime $(M,g)$. 

A spacetime is said to be {\sl future (past) distinguishing} at $x \in M$, if for all $y \neq x$, $I^+(x)
\neq I^+(y) $ ($I^-(x) \neq I^-(y)$). An equivalent formulation can be given in terms of {\sl future (past) locality neighbourhoods}:
\begin{defn}
A neighbourhood \(U\) of \(x\) is said to be a \textsl{future (past) locality neighbourhood} if every future (past) directed causal curve from \(x\) intersects $U$ in a connected set.
\end{defn}
A spacetime is future (past) distinguishing at $x$ if for every neighbourhood $V$ of $x$ there exists
a future(past) locality neighbourhood $U \subset V$ of $x$. If a spacetime is both future
and past distinguishing (FPD) at $x$, then every open set $V \ni x$ contains a
future and past locality neighbourhood of $x$. A future and/or past distinguishing spacetime is one in
which every point is future and/or past distinguishing. Such a spacetime is always causal, but the
converse is not true.

An open set $O$ in $(M,g)$ is {\sl causally convex} if every causal curve between events $x,y \in O$ also
lies in $O$. A spacetime is said to be {\sl strongly causal} at $x$ if $x$ is contained in a
causally convex set whose closure is contained in a simple region. Such a neighbourhood is also
referred to as a {\sl local causality neighbourhood}. Equivalently, every open set $V
\ni x$ contains a neighbourhood $ U$ of $x$ which no causal curve intersects in a disconnected
set. A strongly causal spacetime is one in which all events are strongly causal. Such a spacetime is
always both future and past distinguishing, but again the converse is not always true.  For strongly
causal spacetimes $\msA$ is the manifold topology $\msM$, while it is strictly coarser than $\msM$
when strong causality is violated. Moreover, strong causality is equivalent to $\msA$ being
Hausdorff.

In this work we will emphasise the distinction between manifold topology $\msM$ and {\sl causal
  topology}, the latter being constructed purely from the order relation $\prec$ on $M$. In order to
view the Alexandrov topology as a causal topology, we need to be able to obtain the chronological
relation $\pprec$ from $\prec$. One way of doing this is to first obtain the horismotic 
relation. A natural candidate definition for this relation involves the use of what we shall refer
to as {\sl chain intervals}. 

\begin{defn}
A {\sl chain} $(C,\prec)$ is a totally ordered subset of $(M,\prec)$,
i.e., every pair $u,v \in C$ is such that either $u \prec v $ or $v \prec u$. We say that a causal
interval $[x,y]$ is a {\sl chain interval} if it is a chain. 
\end{defn}

\begin{claim} \label{chain-hor} 
Let $(M,g)$ be a causal spacetime such that for all $p\in M$, every neighbourhood of $p$ contains an incomparable
pair of events. Then $x\rightarrow y$ iff for every $w \in [x,y]$, distinct from $x$ and $y$,
$[x,w]$ and $[w,y]$ are both chain intervals.        
\end{claim} 

\noindent {\bf Proof:} 
If $x\rightarrow y$ then there is a null geodesic from $x$ to $y$. If this null geodesic is unique,
then $[x,y]$ itself is a chain interval and we're done. If it is not unique then (using Propositions
2.19 and 2.20 in \cite{penrose}) for every $z$ with $y\rightarrow z$, we see that $x \pprec z$. For
any $w \in [x,y]$ distinct from $x$ and $y$, therefore, since $x \rightarrow w$ and $w \rightarrow
y$, the null geodesic from $x$ to $w$ is unique as is the null geodesic from $w$ to $y$. Thus,
$[x,w]$ and $[w,y]$ are chain intervals. 

Conversely, for every $w \in [x,y]$, distinct from $x$ and $y$, let  $[x,w]$ and $[w,y]$ both be
chain intervals. Assume that $x \pprec w$. Then $<x,w> \neq \emptyset$. For every $z \in <x,w>$
there is an $O \ni z$ such that $O \subset <x,y>$. By the assumptions of the claim, there exist
$u,v  \in O$ which are incomparable, which is a contradiction. Thus, $x \rightarrow w$ and similarly
$w \rightarrow y$. If $x \pprec y$, then $\exists z$ such that $ x\pprec z \pprec y$. Since $z \in
[x,y]$, this is a contradiction. Thus, $x\rightarrow y$.          
\hfill$\qed$ 

The requirement that every neighbourhood of an event contains an incomparable pair of events seems
general enough to apply to any causal spacetime.  However, one should be careful in attempting a
generalisation. The following example\footnote{ We thank Fay Dowker for this very clarifying
  example.}, though not a counterexample, illustrates the need for care. Consider a $2$ dimensional
spacetime on the cylinder, with the light cones gradually tilting over until there is a single null
geodesic which traverses an $S^1$. Subsequent to this, the light cones then right themselves (see
fig 8.8 of \cite{wald}). One can then get a causal spacetime by removing  a point from the closed null geodesic. Every
$x$ on this null geodesic is however, causally related to every other point in the spacetime -- in
other words, it has no incomparable event! On the other hand, every neighbourhood of $x$ does indeed
contain an incomparable pair.  We will be able to provide an explicit proof that this is a feature of future or
past distinguishing spacetimes, but we do not know if it is true more generally. Indeed, there may be
a broader class of causal spacetimes for which the chain intervals give the horismos relation, but
we will not explore this here.

We  use the language of causal geodesics to define a future directed chain interval from $x$ to
$y$, as the chain interval $[x,y]$ and a past directed chain interval from $x$ to $z$ as the chain
interval $[z,x]$. Similarly, an {\sl open chain interval} is a future and past endless chain $L$
such that for every $x,y \in L$, $[x,y] \subset L$.

\section{Causal but not Strongly Causal Spacetimes} \label{mal} 

\begin{claim} \label{sc} For a future distinguishing spacetime $(M,g)$, every $x \in M$ is
  contained in an arbitrarily small future locality neighbourhood $U \subset M$ such that for every $y \in
  U$ with $y \ssucc x $, $I(x,y)$ is strongly causal. The analogous statement holds in the past distinguishing case. 
\end{claim} 

\noindent {\bf Proof:} Let $(M,g)$ be future distinguishing. For any $y \ssucc x$ which lies in a
future locality neighbourhood $U$ of $x$, $I(x,y)=I(x,y,U)$ (otherwise there is a future directed
causal curve from $x$ to $y$ which leaves $U$ and re-enters). Let $N$ be a simple region containing
$x$ and choose the future locality neighbourhood $U$ of $x$ such that $\overline U \subset
N$. Since $I(x,y)$ is  causally convex  and $\overline{I(x,y)} \subset N$, any $z \in
I(x,y)$ is strongly causal. Similarly for past distinguishing spacetimes. \hfill $\qed$

If $\Delta$ is the set of points in $M$ at which strong causality is violated, then for either past
or future distinguishing spacetimes, the strongly causal region $M-\Delta \neq \emptyset$ and is
moreover open in $\msM$ (see Proposition 4.13 in \cite{penrose}). Thus $\Delta$ is closed in
$M$. We will also employ a minor rephrasing of Theorem 4.31 in \cite{penrose}
\begin{theorem} {\bf Penrose}  \label{penrose} \\
Let $(M,g)$ be a causal spacetime and let strong causality fail at $p\in \Delta \subset M$. Then, 
there is a future and past  endless null geodesic $\Gamma_p$ through $p$ at every point of which strong causality
fails, such that if $u$ and $v$ are any two points of $\Gamma_p$ with $u \prec v$, $u\neq v$, then
$u\pprec x$ and $y\pprec v$  together imply $y \pprec x$.        
\end{theorem}  

We will refer to the above null geodesic $\Gamma_p \subset \Delta$ 
as a {\sl special null geodesic}.  
\begin{claim} \label{unique} 
Let $(M,g)$ be a causal spacetime. Then the special null geodesic $\Gamma_p$ through every $p\in
\Delta \subset M$ is horismotic and unique. 
\end{claim} 
\noindent {\bf Proof:} 
Let $\Gamma_p$ be a special null geodesic through $p\in \Delta \subset M$.  Assume that there exists
a pair $x \prec y $ on $\Gamma_p$ which are not horismotic, i.e., $x \pprec y$. Consider a pair of
events $u,v$ such that $x \pprec u \pprec v \pprec y$. Then by Theorem \ref{penrose}, $v \pprec y$
and $x \pprec u$ implies that $v \pprec u$, which violates causality.  Thus, $\Gamma_p$ is
horismotic.  To show uniqueness, assume that there are two distinct special null geodesics
$\Gamma_p$ and $\Gamma_p'$ through $p \in M$. Let $x \prec p \prec y$ with $x,y \in \Gamma_p$ and
$x' \prec p \prec y'$ with $x',y' \in \Gamma_p'$. Then $x \pprec y'$ and $x' \pprec y$ which from
Theorem \ref{penrose} implies that $x'\pprec y'$, which is again a contradiction since $\Gamma_p'$
is horismotic. \hfill $\qed$

\begin{lemma}\label{regionsc}  
  Let $(M,g)$ be a causal spacetime and let $p$ be a future distinguishing event.  Then there exists
  a neighbourhood $U$ of $p$ such that for any future directed null geodesic segment $\Omega \subset
  U $ from $p$ which is distinct from $\Gamma_p$, $\Omega \cap \Delta =p$, i.e., $\Omega-\{p\}
  \subset M-\Delta$. The analogous time reversed statement holds for $p$ past distinguishing.
\end{lemma} 
\noindent {\bf Proof:} 
Assume to the contrary that no such neighbourhood exists.
Choose $U$ to be a future locality neighbourhood  of $p$, which lies in a simple region, and let $\Omega$ be a future directed null geodesic segment from $p$ such that its intersection with $U$ contains points of $\Delta$ other than $p$.  For such an $s \in \Omega
\cap U$, there exists a special null geodesic $\Gamma_s$ through $s$, which from Claim \ref{unique}
means that $\Gamma_s$ cannot coincide with $\Omega$. For any $r \succ s $ on $\Gamma_s$, $p \pprec
r$. Since $U$ is open, there exits an $r' \in U $ with $r \pprec r' $. From Claim \ref{sc}, since
$<p,r'> \subset U$ is a strongly causal region, this is a 
contradiction. Similarly for a past distinguishing event.\hfill $\qed$

\begin{figure}[ht]  
\centering 
\includegraphics[width=5cm, height=5cm]{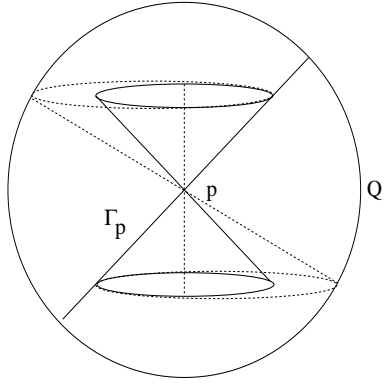}
\label{scregion}
\caption{{\small The regions $J^\pm(p,Q)$ for $Q$ a future and past locality neighbourhood of $p$
    are, excluding $\Gamma_p$, strongly causal. Since the regions of strong causality are open, we depict
    the strongly causal regions in $Q$ as the  interiors of two widened cones from $p$ which contain the sets
    $J^\pm(p,Q)$, and intersect them only along $\Gamma_p$. }}
\end{figure}

In particular, this means that every neighbourhood of a future distinguishing event $p$ in a causal spacetime 
contains a future locality neighbourhood $Q \ni p$ such that $J^+(p,Q)\backslash \Gamma_p$ lies in $M -
\Delta$, where $\Gamma_p$ is the special null geodesic through $p$. We illustrate this in Fig
\ref{scregion}. 

Let us now cast these results into the language of chain intervals as promised. In order to use
chain intervals interchangeably with null geodesic segments, we need to satisfy the conditions of
Claim \ref{chain-hor}. Equivalently, it suffices to prove that 

\begin{claim} \label{forpd}
 For a future or past distinguishing spacetime, if $[x,y]$ is a chain interval then
  $x\rightarrow  y$. 
\end{claim} {\bf Proof:} Assume otherwise, i.e., let $x \pprec y$, and let $O$ be an open set
contained in $<x,y> \neq \emptyset$. From Lemma \ref{regionsc} every future or past locality neighbourhood $U$ of  $p \in M$
intersects the strongly causal region $M-\Delta$ non-trivially.  For $p \in O$, choose $U \subset
O$  to be such a neighbourhood of $p$, and let $r$ be a strongly causal point in $U$.  Now, every causally convex
neighbourhood $W$ of $r$ contains an $s$ which is incomparable to it.  If  $W$ is chosen to be a
subset of $U$ then $r,s$ are an incomparable pair in $<x,y> \subset[x,y]$, which is a contradiction. \hfill $\qed$ 

This allows us to give a local characterisation of strong causality: 

\begin{lemma}\label{scchain} A future or past distinguishing spacetime is strongly causal at $p$ iff
  every null geodesic segment containing $p$ in its interior contains a chain interval with $p$ in its interior.
\end{lemma} 

\noindent {\bf Proof:} Let $p$ be strongly causal and $U$ a local causality neighbourhood of $p$.
Every null geodesic segment in $U$ through $p$ is horismotic in $U$ and is
therefore a chain interval. Conversely, if $p$ is not strongly causal, then by Theorem \ref{penrose}
there is a special null geodesic $\Gamma_p$ through $p$. Let $\Omega$ be a null
geodesic through $p$ distinct from $\Gamma_p$ and let $a, b \in \Gamma_p$ with $a \prec p
\prec b$. Then for any $x, y \in \Omega$ such that $x \prec p \prec y$, $x \pprec b$ and $a \pprec
y$. By Theorem \ref{penrose} this means that  $ x \pprec y$. Since the spacetime is future or past
distinguishing, $<x,y>$ contains incomparable pairs of events and hence $[x,y]$ is
not a chain interval. Since this is true for any pair $x,y$, we see that no null geodesic through
$p$ which is distinct from $\Gamma_p$ contains a chain interval with $p$ in its interior.
\hfill $\qed$

Using Claim \ref{unique} this means that 
\begin{claim}\label{cint} 
If strong causality fails at a point $p$ in a future or past distinguishing spacetime then the only chain interval which
contains $p$ in its interior lies in $\Gamma_p$.   
\end{claim} 

On the other hand, 
\begin{claim} \label{fpdchains} 
  If $p$ is a future distinguishing point in a causal spacetime, then any future directed null geodesic from $p$
  contains a chain interval $[p,w]$. The analogous time-reversed statement holds for $p$ past distinguishing.
\end{claim} 

{\bf Proof:} Let $p$ be future distinguishing and let $U$ be a future locality neighbourhood of $p$
which lies in a simple region.  Let $N$
be a null geodesic from $p$ and $z \in N \cap U$. Since $J^+(p,U)=J^+(p) \cap U$ this means that
$p\rightarrow z$. Then for any $w \in [p,z]$, $[p,w]$ is a chain interval by the above claim.
\hfill $\qed$

It is important to know whether causal properties of spacetimes are preserved under a causal
bijection. In \cite{levichev} it was shown that for future and past distinguishing spacetimes, a
causal bijection is also a chronological bijection. Using the results above, we can extend this to
the slightly more general statement 

\begin{claim} \label{extlev}
If $f:(M_1,g_1) \rightarrow (M_2,g_2)$ is a causal bijection between two spacetimes both of which
are either future distinguishing or past distinguishing, then $f$ is also a chronological bijection.   
\end{claim} 
\noindent {\bf Proof:} First note that a causal bijection preserves chain intervals.  Let
$x_1,y_1 \in M_1$ such that $x_1 \rightarrow_1  y_1$. Then by Claims \ref{chain-hor} and \ref{forpd}
for every $w_1 \in [x_1,y_1]_1 $,  $[x_1,w_1]_1$ and $[w_1,y_1]_1$ are chain intervals. Now, $f(x_1 )
\prec_2 f(y_1)$. Assume that $f(x_1) \pprec_2 f(y_1)$. Then by the proof of Claim \ref{forpd} we see
that  $[f(x_1), f(y_1)]_2$ is not a chain interval. Moreover, there exists a $w_2 \in <f(x_1), f(y_1)>_2
\neq \emptyset$ such that $f(x_1) \pprec_2 w_2 \pprec_2 f(y_1)$, so that $[f(x_1),w_2]_2$ and
$[w_2,f(y_1)]_2$ are not chain intervals. Since $f^{-1}(w_2) \in [x_1,y_1]_1$ this is a
contradiction.  Thus if $x_1 \rightarrow_1 y_1$ then $f(x_1) \rightarrow_2 f(y_1)$. Conversely,
by the same argument, if $x_1 \pprec_1 y_1$ then $f(x_1) \pprec_2 f(y_1)$.    
\hfill $\qed$ 



\begin{claim}
  Let $f$ be a causal bijection between two spacetimes $(M_1,g_1)$ and $(M_2,g_2)$ both of which are
  either future or past distinguishing. Then if $p \in M_1$ is strongly causal in $(M_1,g_1)$ 
  $f(p) \in M_2$ is strongly causal in $(M_2,g_2)$.
\end{claim}  

\noindent {\bf Proof:} 
Assume otherwise, i.e., that $f(p)$ is not strongly causal.  At  $p$ we can find  two distinct chain intervals  $[x,y]$ and
$[u,v]$ that contain $p$. This means that $x \pprec v$ and $u \pprec y$.   Since $[f(x),f(y)]$
and $[f(u),f(v)]$ are also chain intervals, and they  contain $f(p)$, they must lie in
the special null geodesic $\Gamma_{f(p)}$ (by Claim \ref{cint}). But since $f$ also preserves
chronology by Claim \ref{extlev},  $f(x) \pprec f(v) $ which 
is a contradiction, since  $f(x), f(v) \in \Gamma_p$.  \hfill $\qed$ 

We are now in a position to prove Proposition \ref{claim-two}: 

\noindent {\bf Proof of Proposition \ref{claim-two}:} 
Wlog let the two spacetimes $(M_1,g_1)$ and $(M_2,g_2)$ both be future distinguishing. Every  $p \in
M_1$ is contained in a future locality neighbourhood $U$ such that $<p,x>$ is strongly causal for
every $x \in U$. Since $<p,x>$ is causally convex and is open in the manifold topology,
it can also be topologized with Alexandrov intervals in $<p,x>$, so that $\msA_1|_{<p,x> }=
  \msM_1|_{<p,x>}$ because of strong causality. Similarly, since $f(<p,x>)=<f(p),f(x)>$ is also
  strongly causal $\msA_2|_{<p,x> }= \msM_2|_{<f(p),f(x)>}$. However, since $f$ is a homeomorphism from
    $\msA_1$ to $\msA_2$ it is also a homeomorphism between $\msM_1|_{<p,x>}$
    and $ \msM_2|_{<f(p),f(x)>} $ which implies that   $M_1$ and $M_2$ have the same dimension.
\hfill $\qed$ 

A generalisation of the MHKML theorem is then immediate
\begin{corollary} {\bf Extension of MHKML:} \\
  If a causal bijection $f$ exists between two spacetimes of dimensions $n_1, n_2 >2$ which are both
  future and past distinguishing, then $n_1=n_2$ and the spacetimes are conformally isometric.  
\end{corollary}

It is important to note that while there is no restriction to $n>2$ in Proposition \ref{claim-two},
such a restriction is crucial to the  Hawking-King-McCarthy result and hence the above Corollary.  

\section{ A Topology based on  Convergence of Chain Intervals}\label{newtop}

The fact that causal bijections encode manifold dimension and topology so generally is interesting and begs the
question of whether there exists a causal topology which is equivalent to the manifold topology
when the spacetime fails to be strongly causal. In this section we present a new topology for future
or past distinguishing spacetimes, using an Alexandrov convergence of chain intervals which we term
$\msN$-convergence . In general, this convergence criterion is not equivalent to manifold
convergence but for future and past distinguishing spacetimes, for which the region of strong
causality violation is locally achronal, we can demonstrate that $\msN$-convergence is the same as
manifold convergence.

Just as the convergence of causal curves is defined with respect to the manifold topology, one can
also define the convergence of chain-intervals with respect to a causal topology like the
Alexandrov topology.  A sequence of chain-intervals $\{\Omega_i = [p_i,q_i]\}$ will be said to
Alexandrov converge to an event $x \in M$ if for every $\msA$ open neighbourhood $A$ of $x$ there
exists an $N$ such that for all $i> N$, $\Omega_i \cap A$ is non-empty. In what follows we use the
Alexandrov convergence of chain intervals to define a new convergence condition for the end points
$p_i$ or $q_i$ of chain intervals.

\begin{defn}
Let $(M,g)$ be a causal spacetime and let $\Delta$ be the region of
strong causality violation. A sequence $\{p_i\} $ is said to {\sl future $\msN$-converge} to $p$ if
there exists a future directed chain interval $\Omega^+ =[p,q]$ from $p$ with $\Omega^+ - \{ p\} \subset
M-\Delta$ and a sequence of future directed {\it non-intersecting} chain intervals
$\{\Omega_i^+=[p_i, q_i]\}$ from $p_i$ such that every point in $\Omega^+-\{p\}$ is an
Alexandrov convergence point of the sequence $\{\Omega_i^+\}$ and, moreover no subsequence of
$\{\Omega_i^+\}$ has any other convergence points off $\Omega^+$ in $M-\Delta$. Past $\msN$-convergence is similarly defined.
\end{defn}

This definition is inspired by a construction used in Malament's paper \cite{malament} which employs
null geodesics instead of chain intervals. For completeness and later comparison, we define the 
null geodesic version of $\msN$-convergence as:

\begin{defn} 
Let $(M,g)$ be a causal spacetime and let $\Delta$ be the region of
strong causality violation.  A sequence $\{p_i\} $ is said to {\sl future geodesic-$\msN$-converge}
to $p$ if there exists a future directed null geodesic segment $\gamma^+$ from $p$ with $\gamma^+-
\{ p\} \in M-\Delta$ and a sequence of future directed non-intersecting null geodesic segments
$\{\gamma_i^+\}$ from $p_i$ such that every point in $\gamma^+-\{ p\}$ is an Alexandrov convergence
point of the sequence $\{\gamma_i^+\}$ and, moreover  no subsequence of $\{\gamma_i^+\}$ has any other convergence points off
$\gamma^+$ in $M -\Delta$. Past geodesic $\msN$-convergence is similarly defined.
\end{defn}

We will find it useful to first show the following, straightforward, result and then employ it to
deal with the more complicated case of future and past distinguishing spacetimes.  
\begin{lemma}\label{mink}  
Future or past $\msN$-convergence are equivalent to manifold convergence in Minkowski spacetime. 
\end{lemma} 

\noindent {\bf Proof:} 
Let $\{ p_i\} \xrightarrow{\msM} p$. Any future or past directed null geodesic segment from $p$ is a
chain interval and since $\Delta = \emptyset$, it lies in $M-\Delta$. Let $O$ be an open
neighbourhood of $p$ and let $\Omega^+$ be a future directed null geodesic segment from $p$ which is
future inextendible in $O$. Then there exists a non-contracting or expanding future directed null
geodesic congruence $\Omega_\alpha^+$ in $O$ which contains $\Omega^+$, generated by a null-vector
field (strictly an equivalence class of null-vector fields) $\xi^a$. If $\Omega$ is the (unique)
past completion of $\Omega^+$ in $O$, then $\Omega_\alpha$ is the corresponding null geodesic
congruence that continues $\Omega_\alpha^+$ in the past. For the sub-sequence of the $\{ p_i\}$'s
which lie in $O$, let $ \Omega_i^+$ be the future directed null-geodesic in this congruence
from $p_i$ which is future inextendible in $O$.  $\Omega_i$  are chain intervals since the spacetime
is Minkowski.

Let $q \in \Omega^+$ such that $q$ is not an Alexandrov (and hence manifold) limit point of the
sequence $\{ \Omega_i^+\}$.  Then there exists a neighbourhood $U \subset O$ of $q$ which does not
intersect any of the $\{ \Omega_i^+\}$.  Now, $\xi^a$ generates a one parameter family of
diffeomorphisms and hence $U$ defines a collar neighbourhood $\cT_\xi(U)$ of $\Omega$. Since $U \cap
\Omega_i^+=\emptyset$ for all $i$ this means that none of the $\Omega^+_i$ can enter $\cT_\xi(U)$
which means that there is a neighbourhood $O' \subset \cT_\xi(U)$ of $p$ which intersects none of
the $\Omega_i$ which is a contradiction.  Therefore every $q \in \Omega^+$ is a limit point of $\{
\Omega_i^+\}$.  Moreover, since the $\Omega_i^+$ and $\Omega^+$ are future inextendible in $O$ all
convergence points of $\{\Omega_i\}$ lie on $\Omega^+$. Therefore the $\{p_i\}$ future $\msN$-converge
to $p$. A similar argument shows that the $\{p_i\}$  also past $\msN$-converge to
$p$. 

Conversely, let $\{ p_i\}$ future $\msN$-converge to $p$. Then there exists a future directed chain
interval $\Omega_+$ from $p$ and a sequence of future chain intervals $\Omega_i^+$ from $p_i$ such
that every point on $\Omega^+-\{p\}$ is an Alexandrov convergence point of $\{\Omega_i^+\}$ with no
other convergence points besides $p$.  If $\{ p_i\}$ does not converge in $\msM $ to $p$ then there
exists a neighbourhood $U$ of $p$ which does not contain any of the $p_i$. On the other hand since
every point on $\Omega^+\cap U - \{p\}$ is an Alexandrov and hence a manifold convergence point for
$\{ \Omega_i^+\}$, there exists an $N$ such that for all $i>N$, $\Omega_i^+ \cap U \neq
\emptyset$. Thus each $\Omega_i^+$ must enter and then leave $U$ to reach $p_i$ to the
past. However, in $U$ since each $\Omega_i^+$ is also a null-geodesic segment, it is uniquely
defined and hence it must also have points of Alexandrov
convergence on the past-extension $\Omega^-$ of $\Omega^+$ in $U$, which is a contradiction since
$\Delta=\emptyset$.   
\hfill $\qed$

For a generic spacetime some elements of the above proof are still valid, as long as we restrict to
a simple region $O$ around $p$. However, since the Alexandrov topology is used in defining
$\msN$-convergence, rather than the manifold topology, care has to be exercised in the
generalisation.  In a strongly causal spacetime, for example, since $\msA \sim \msM$, this
distinction is no longer important and the existence of arbitrarily small causally convex
neighbourhoods of $p$ implies that the entire proof of Lemma \ref{mink} can be reproduced for this
case.

\begin{figure}[ht] \label{doublesheet} 
\centering \resizebox{2in}{!}{\includegraphics{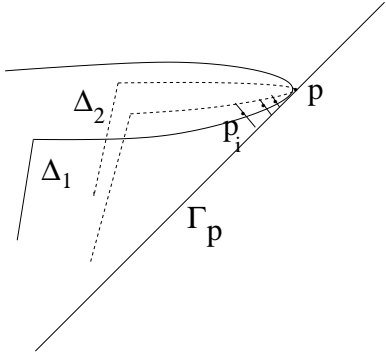}}
\vspace{0.5cm}
\caption{{\small The strong causality violating region around $p \in \Delta$ can have a complicated
    structure. Here, the regions $\Delta_1$ and $\Delta_2$ intersect only on  $\Gamma_p$. They are
    each 
    spacelike with respect to $\Gamma_p$, but there is no neighbourhood of $p$ in which $\Delta$ is
    achronal. A sequence $\{p_i\}$ which manifold converges to $p$ can be  trapped in between
    $\Delta_1$ and $\Delta_2$ as shown. Even though the $p_i$ can eventually lie off $\Delta$, chain
    intervals from and to $p_i$ can manifold converge only to $p$ since they are also trapped in
    this region.}}
\end{figure}

When strong causality is violated, however, much more caution is required. From Lemma \ref{regionsc}
we see that if the spacetime is future distinguishing at $p$ it admits a future directed chain
interval $\Omega^+$ from $p$ with $ \Omega^+ - \{ p\}$ in $M-\Delta$ and similarly if it is past
distinguishing at $p$ it admits a past directed chain interval $\Omega^-$ from $p$ with $\Omega^--
\{ p\} \in M-\Delta$.  However the converses are not always possible.  Thus, in order to be able to
equate manifold convergence to either future or past $\msN$-convergence, one needs the spacetime to
be both future and past distinguishing. However, even this is not quite enough.  Although a {\it
  null geodesic} congruence can be constructed through the $p_i$'s which Alexandrov converge to all
points on $\Omega^+-\{ p\}$ these do not necessarily give rise to a sequence of chain intervals that
Alexandrov converge to all points on $\Omega^+-\{ p\}$.  This is because even though $\Delta$ is
locally achronal with respect to $p$ it need not itself be locally achronal -- i.e.,  there may exist
no open neighbourhood $U \ni p$ in the manifold topology such that $\Delta \cap U$ is
achronal. Hence the null geodesic congruence through $p_i$ could, for all $i$, intersect $\Delta
$ both in the future and the past. This means that the corresponding chain intervals that one can construct are
trapped between different ``leaves'' of $\Delta$ and hence manifold convergence would not 
imply  $\msN$-convergence for such ``trapped'' sequences. Figure \ref{doublesheet} illustrates the
problem. 

\begin{defn} 
The region of strong causality violation $\Delta $ in a causal spacetime $(M,g)$ is said
to be {\sl locally manifold achronal} if for every $p \in \Delta$ there exists an open neighbourhood $U \ni p$ in the
manifold topology such that $\Delta \cap U$ is achronal. It is said to be {\sl locally Alexandrov
  achronal}  if $U$ is  required instead to be open in
the Alexandrov topology. 
\end{defn}

Since $\msA$ is strictly coarser than $\msM$ for spacetimes that are not strongly causal, and in
particular for those Alexandrov sets which contain events in $\Delta$, the requirement of local
Alexandrov achronality is stronger than the manifold version. Thus, although one might prefer using
the former because it is intrinsically causal, it is from the spacetime perspective, more
restrictive than necessary to establish the equivalence of $\msN$-convergence to manifold
convergence. For this purpose, we will use only the manifold version of local achronality for
$\Delta$, which is then also true for  Alexandrov local achronality.

In order to deal with all manner of manifold converging sequences, either past or future
$\msN$-convergence is required, depending on whether the sequence lies in the causal
future of a local patch of $\Delta$ or to its causal past. 

\begin{defn}
$\{ p_i\}$ is said to $\msN$ converge to $p$ if it either future {\it or} past
$\msN$-converges to $p$.
\end{defn}

\begin{prop} \label{equivconv}
  Let $(M,g)$ be a future and past distinguishing spacetime. Then $\msN$-convergence is equivalent
  to manifold convergence everywhere in $M$ if either $\Delta= \emptyset$, or it is locally
  manifold achronal.
\end{prop}

\noindent {\bf Proof:} If the limit point $p \in M- \Delta$, then the proof is similar to that of
lemma 5; so in what follows we will assume $p \in \Delta$. Let $\{ p_i\} \xrightarrow{\msM} p$. Let
$U$ be a future and past locality neighbourhood of $p$ which is contained in a simple region
$O$. Let $\Omega$ be a null geodesic through $p$ with $\Omega - \{ p\} \subset M-\Delta$ which is
both past and future inextendible in $O$. If $\Omega^+$ and $\Omega^-$ are the segments of $\Omega$
that are to the causal future and the causal past of $p$ respectively, then for any $q \in \Omega^+$
and $r \in \Omega^-$, $L^+_q\equiv [p,q]$ and $L^-_r \equiv [r,p]$ are both chain intervals. In
particular $L^+\equiv \Omega^+$ and $L^- \equiv \Omega^-$ are both ``open'' chain intervals in $O$
(i.e. future and past endless in $O$, respectively, as defined in Section \ref{prelims}).

We further assume that $O$ is contained in a neighbourhood $Q$ of $p$ in which $\Delta$ is achronal.
Let $\Omega_i$ be null geodesics through the $p_i$ which belong to a locally non-singular null
geodesic congruence constructed as in Lemma \ref{mink}. Let $\Omega_i^+$ and $\Omega_i^-$ be the
future and past segments of $\Omega_i$ respectively. Using the exponential map, we see that every $q
\in \Omega^+-\{p\}$ is a manifold convergence point of $\{ \Omega_i^+ \}$ and since $\Omega^+-p \subset
M-\Delta$ it is also an Alexandrov convergence point. In particular, there is no other manifold and
hence Alexandrov convergence point off $\Omega^+$.   
   
In order to construct chain intervals from $\Omega_i^\pm$, one has to see whether they intersect
$\Delta$ or not. Since $\Omega - \{p\} \subset M - \Delta$ and every $q \in \Omega-\{p\}$ is an Alexandrov
convergence point of $\{ \Omega_i \}$, eventually $\Omega_i \not \subset \Delta$. If $\Omega_i \cap
\Delta $ is either empty or just $p_i$ then both $\Omega_i^+$ and $\Omega_i^-$ are open chain
intervals. 
Since $O$ is not a subset of  $\Delta$, $\Omega$ can at worst intersect $\Delta$ at two
disjoint points $p_1 $ and $p_2$. However, since $\Delta$ is achronal in $O$ this leads to a
contradiction: since $p_{1,2} \in \Delta$ $\,\, \exists r_1 \prec p_1$ with $r_1 \in \Gamma_{p_1}$,
and since $p_2 \nin \Gamma_{p_1}$, $\Rightarrow r_1 \pprec p_2 $ which is a contradiction.



Let each $\Omega_i$ intersect $\Delta \cap O$ at $q_i$ or nowhere at all. If $p_i \succ q_i$ then
$L_i^+ \equiv \Omega_i^+$ is an open chain interval and if $p_i \prec q_i$ then
$L_i^- \equiv \Omega_i^-$ is an open chain interval. Now, for every sequence $\{p_i\}$ one can
extract a subsequence $\{p_i' \}$ such that either all the $\{\Omega_i^+\}$ or all the $\{
\Omega_i^-\}$  are open chain intervals. Thus, $\{p_i \} \xrightarrow{\msN} p$.

The converse argument is similar to that in Lemma \ref{mink}. Let $\{p_i \} \xrightarrow{\msN} p$ and
assume wlog that this is future $\msN$-convergent. Namely, there exists a sequence of future chain
intervals $\{ L_i^+\}$ from $p_i$ which Alexandrov converges off $p$ to a chain interval
$L^+$ from $p$ where $L^+-\{p \} \subset M -\Delta$, with no other convergence points in $M-\Delta$.  
Now assume contrary to the assertion, that $\{p_i\}$ does not $\msM$-converge to $p$. Then there
exists a simple region $O \ni p$ which eventually contains none of the $p_i$. 
Then $L^+$ and $L_i^+$ correspond to future directed null geodesic segments from $p$ and $p_i$
respectively, and have  unique completions in $O$ both to the past and the future. If $L^-$ is the
unique null geodesic past extension of $L^+$ in $O$, then it cannot lie in $\Delta$ by Claim
\ref{unique}. The arguments of Lemma \ref{mink} can then be reproduced to show that $\{ L_i^+\}$
also Alexandrov converges to $L^-$ which is a contradiction.  Thus, $\{ p_i\} \xrightarrow{\msM}
p$.  \hfill $\qed$

From the above proofs it is clear that future or past geodesic $\msN$-convergence, which uses null
geodesic segments instead of chain intervals, is equivalent to manifold convergence for {\it all} future and
past distinguishing spacetimes without the further assumption that $\Delta$ be  locally achronal.  

\begin{defn}
A set $O$ is $\msN$-\textsl{open} if every sequence of points in its complement $O^c$, which $\msN$-converges, has its limit
point in $O^c$.
\end{defn}

Let $\msN$ denote the collection of $\msN$-open sets.

\begin{lemma} 
The collection of sets $\msN$ forms a topology on $M$. 
\end{lemma} 

\noindent {\bf Proof:} To show that $\msN$ forms a topology on $M$, we need to prove that it
satisfies the three properties of a topology. (i) The proof of $M,\emptyset \in\msN$ is
trivial. (ii) Next let $U_{\alpha}$ be sets in $\msN$, where $\alpha$ belongs to some
indexing set. To show that $\bigcup_{\alpha}U_{\alpha}\in \msN$, note that if an $\msN$-convergent sequence lies outside $\bigcup_{\alpha}U_{\alpha}$, then it
must lie outside each of the $U_{\alpha}$. Since $U_{\alpha}\in \msN$ for all
$\alpha$, the limit point of the sequence must also lie outside $U_{\alpha}$ for all
$\alpha$. Thus, the limit point is not contained in $\bigcup_{\alpha}U_{\alpha}$, which proves
that $\bigcup_{\alpha}U_{\alpha}\in \msN$. (iii) To show that
$\bigcap_{i=1}^nU_{i}\in \msN$, let $\{x_i\}$ be an $\msN$-convergent sequence in $M-\bigcap_{i=1}^nU_{i}=\bigcup_{i=1}^n(M-U_{i})$. Since the $(M-U_{i})$'s
are finite in number, at least one of them, say $(M-U_{k})$ should contain a subsequence $\{y_j\}$
of $\{x_i\}$. It is clear from the definition of $\msN$ convergence that every subsequence of an $\msN$-convergent sequence also $\msN$-converges to the same limit point. 
But since $U_{k}\in \msN$, the limit point of $\{y_j\}$ must lie in $(M-U_{k})$. Thus, the limit point of
$\{x_i\}$ is in $\bigcup_{i=1}^n(M-U_{i})=M-\bigcap_{i=1}^nU_{i}$, and $\bigcap_{i=1}^nU_{i}\in
\msN$.  \hfill $\qed$

Since $\msM$ is metrizable, and therefore first countable, lemma \ref{equivconv} implies that for an
FPD spacetime $(M,g)$ in which the strong causality violating set satisfies local manifold achronality, the two topologies $\msM$ and $\msN$ are equivalent. To see this, consider an $\msN$ open set $O$ in $M$. If $O$ were not $\msM$ open, then from the first countability of $\msM$, there exists a sequence $\{x_i\}$ entirely contained in $O^c$ which $\msM$-converges to a point $x\in O$. But from lemma \ref{equivconv}, this would mean that $\{x_i\}$ $\msN$-converges to $x$, contradicting the fact that $O$ is $\msN$ open. Similarly, if $U$ is an $\msM$-open set, then it follows that $U$ is also $\msN$ open, because otherwise the definition of $\msN$ would imply the existence of an $\msN$-convergent, and thus $\msM$-convergent sequence contained entirely in $U^c$, with it's limit point in $U$, a contradiction. 

 Although $\msN$ is a causal topology, we do not yet have a useful basis representation unlike the
 Alexandrov intervals $<x,y>$ for $\msA$ and the Fullwood ``double'' intervals $<x,y,z>$ for $\msF$
 described in the following section. We end this section with a discussion on the type of
 difficulties one encounters in trying to construct a manifold local causal basis. We first explore in
 more detail the role played by chain intervals in a future or past distinguishing spacetime.

 Let $(M,g)$ be strongly causal at $p$. If $L$ is the set of all chain intervals that contain $p$ in
 its interior, define the equivalence relation $\sim $ as follows. For every $l_1, l_2 \in L$ $l_1
 \sim l_2$ if $l_1\cap l_2 -\{p\} \neq \emptyset$. Since every null geodesic with $p$ in its
 interior contains a chain interval with $p$ in its interior (Lemma \ref{scchain}) there exists a
 bijection $\mu: [L] \rightarrow S^{n-2}$, where $[L]$ are the equivalence classes of chains under
 $\sim$ and $S^{n-2}$ represents the set of either future or past null-directions from $p$.
 Similarly, if the spacetime is future distinguishing at $p$ and if $L^+$ is the set of chain
 intervals of the form $[p,q]$, then again $\sim$ can be used to determine an equivalence between
 chains. Using Claim \ref{cint}, we see that $\mu_+: [L^+] \rightarrow S^{n-2}$ is a bijection where
 $S^{n-2}$ is the set of future null directions from $p$.  $L^-$ is similarly defined for a past
 distinguishing point $p$ with $\mu_-: [L^-]\rightarrow S^{n-2}$ and $S^{n-2}$ is now the set of
 past null directions from $p$.  Requiring a spacetime to be both future and past distinguishing
 means that one has a purely local causal definition of the future and past light cones emanating
 from each point in the spacetime.  In particular, for every $p \in \Delta$ which is both future and
 past distinguishing there is an $S^{n-2} - \{\mathrm{point}\}$  worth of (equivalence classes of) chain intervals
 from $p$ which lie in the strongly causal region $M-\Delta$ (see Figure \ref{scregion}).
 Conversely, without future or past distinguishability, there seems to be no local 
 definition of future or past light cones, respectively. 

Lemma \ref{scchain} moreover allows us to characterise the null geodesics in a strongly causal
spacetime in a natural way. Let $(C,\prec)$ be a chain or totally ordered subset of $(M,\prec)$.  If
every $x \in C$ lies in the interior of a chain interval which itself lies in $C$ then we say that
$(C,\prec)$ is a {\sl locally causally rigid} chain (LCRC). Such a chain is also suitably dense and
corresponds to a null geodesic in the spacetime. In a strongly causal spacetime, this provides a
purely causal description of an arbitrary null geodesic. If on the other hand a chain contains an
event $x$ at which strong causality is violated, then it cannot be locally causally rigid past this
point, unless it is itself the special null geodesic containing $x$. Thus, if an LCRC contains an
$x \in M-\Delta$, then it must lie in $M-\Delta$. We can at best attach end points to an LCRC which
are not required to be locally rigid themselves and hence can lie on $\Delta$. Let $C$ be an LCRC
with a future end point $x \in \Delta$, such that $C \not\subseteq \Gamma_p$. In a future and past
distinguishing spacetime, it is possible to begin a new future directed LCRC $C'$ from $x$, but
there is no natural causal choice of $C'$ that ensures that $C\cup C'$ is a null geodesic. This
directional ``floppiness'' at $x$ is what makes the construction of a basis for the topology $\msN$
particularly difficult.

It is useful to examine the structure of Alexandrov intervals in spacetimes in which
strong causality is violated.  We notice that 
\begin{claim} 
For a causal spacetime $(M,g)$, if $<x,y>$ contains a $p \in \Delta$, then $\Gamma_p \subset \overline{<x,y>}$.   
\end{claim}  

\noindent {\bf Proof:} For any $q \succ p$, $q \in \Gamma_p$, $z \pprec q \Rightarrow z \pprec
y$. Moreover, since $ x \pprec p \prec q \Rightarrow x \pprec q$, every $z \in <x,q>$ also lies
in $<x,y>$. Similarly, for $r \prec p$ $r \in \Gamma_p$, $w \succ r \Rightarrow w \succ x$ and
since $r \pprec y$, every $w \in <r,y>$ belongs to $<x,y>$. Since this is true for every $q> p$
on $\Gamma_p$ and every $r<p$ on $\Gamma_p$, we have the desired result.  \hfill $\qed$ 

\begin{figure}[ht]
\centering \resizebox{4in}{!}{\includegraphics{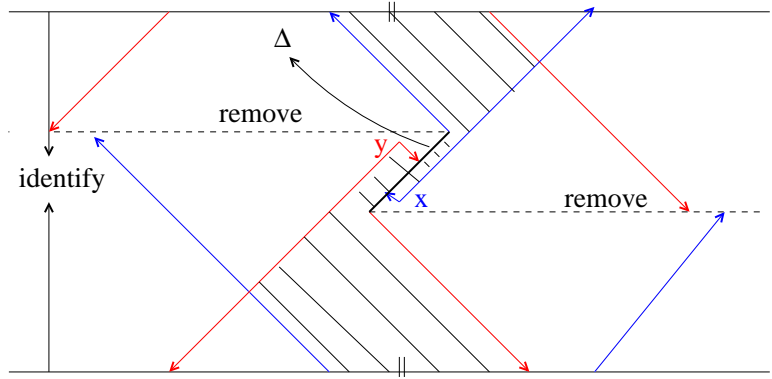}}
\vspace{0.5cm}
\caption{{\small In a spacetime which is distinguishing but not strongly causal the Alexandrov
    interval can have a very non-local character. In the spacetime depicted above, the thickened
    black line is the only special null geodesic in $M$ and is all of $\Delta$. The Alexandrov
    interval $<x,y>$ for two events which straddle $\Delta$ is given by the shaded region. }}\label{alexint}
\end{figure}

At this point it is useful to make use of the specific 2 dimensional example of Figure
\ref{alexint}.  This spacetime has a single special null geodesic $\Gamma$ and for pairs $x \pprec
y$ which straddle $\Gamma$ the Alexandrov interval $<x,y>$ ``spreads'' across the spacetime. This
example makes explicit the fact that $\msA$ is strictly coarser than $\msM$ in such spacetimes.
Thus, the Alexandrov interval is not sufficiently local to be useful.  On the other hand, if  $O$ is manifold open
  then the  restricted Alexandrov interval $<x,y>_O \subset
O$ if $x,y \in O$. This restricted interval  is appropriately manifold local and one might hope to
find a purely causal way of defining such a set. In our 2 dimensional example, we can define
this set via its boundary. For every $x\in M$, there are exactly two forward directed and two
backward directed null geodesics and hence (classes of) chain intervals. One of the future directed
pairs of chains starting from $x$, $f_x^+$ hits $\Gamma$ at some $z$ before it can hit a past
directed chain from $y$. Across $z$, it is no longer a chain interval. Thus, at $z$, we have two
future directions again to take, $f_z^+$ and $f_z^-$. The latter choice lies along $\Gamma$ and we
may reject it and instead take the union of chains $f_x^+ \cup f_z^+$. This intersects a past
directed chain from $y$, $p_y^+$ at some $r$. A similar construction from the past directed chains
from $y$ carves out the boundary of a region with the desired local properties.

Even if such a construction were universally possible, how would one define the interior region
purely causally? This is surprisingly difficult even for the simple spacetime under consideration,
without the added complication of higher dimensions. However, by suitably ``carving out''
sufficiently manifold local sets, using locally defined sequences of chain intervals it may be
possible to make further progress on this question.

\section{Other Causal Topologies}\label{other} 
         
In \cite{fullwood} a new topology $\msF$ was constructed from a basis of sets obtained by taking the
union of two Alexandrov intervals $<x,y,z>\equiv <x,y> \cup <y,z> \cup y$. These sets are not open
in the manifold topology since they include the intermediate point $y$. For a $y \in \Delta$,
$<x,y,z>$ is an $\msF$ open neighbourhood of $y$ and for appropriate choices of $x$ and $z$, either
$<x,y> \subset M-\Delta$ and or $<y,z> \subset M-\Delta$ depending on whether the spacetime is
future or past distinguishing. Thus, by the arguments in Section \ref{mal}, $\msF$ also contains
information about the spacetime dimension. $\msF$ is Hausdorff iff the spacetime is future and past
distinguishing and is moreover, strictly finer than $\msM$ \cite{fullwood}.

It was shown in \cite{fullwood}, that $\msF$ can also be obtained via a causal convergence criterion
on timelike sequences of events. As we will show below, a slight generalisation of this definition
to include all monotonic \textsl{causal} sequences, gives rise to yet another causal topology which
we call $\msP$. We will show however, that $\msP$ is strictly coarser than $\msF$ and also strictly
finer than $\msM$. Some of the following definitions and results have also been considered in a somewhat different framework in \cite{Harris1,Harris2}.

\begin{defn} 
A sequence $\{ p_i\}$ is said  to be \textsl{future directed monotonic timelike} if for $i<j$ $p_i
\pprec p_j$, and is \textsl{past directed monotonic timelike} if for $i<j$ $p_i\ssucc
p_j$. We can similarly define future and past directed monotonic \textsl{causal} sequences. 
\end{defn}

In \cite{fullwood} causal convergence is defined as follows. 

\begin{defn}
A monotonic causal sequence $\{p_j\}$ is said to \textsl{causally
  converge} to $p$ as $j\rightarrow \infty$ if either (a) $I^{-}(p) =
\bigcup_{i}I^{-}(p_i)$ or (b) $I^{+}(p) = \bigcup_{i}I^{+}(p_i)$.
\end{defn}



Before discussing  the relation between these various topologies, we pause to
outline some properties of monotonic causal sequences and the $\msP$-topology.

\begin{lemma}  \label{subseq}
A future or past monotonic causal sequence $\{x_i\}$ causally converges to $x$, iff
every (infinite) subsequence of $\{x_i\}$ also causally converges to $x$.
\end{lemma} 
\noindent \textbf{Proof}: First, let $\{x_i\}$ be a future-directed monotonic causal sequence which causally
converges to $x$ and let $\{y_j\}$ be a subsequence of $\{x_i\}$. Thus
$\bigcup_{j}I^{-}(y_j)\subset  
\bigcup_iI^{-}(x_i) = I^-(x)$. For any  $z\in I^{-}(x)$, there exists a $k$ such that for all $n\geq
k$,    $z\in I^-(x_n)$. Since $\{x_i \}$ is future directed monotonic and  $\{y_j\}$ is an infinite
subsequence, there exists $l$ such that $y_l=x_n$ for some $n\geq k$ so that $I^-(x) \subset 
\bigcup_j I^-(y_j)$ so that  $I^-(x) = \bigcup_j I^-(y_j)$. 

Conversely, let $\{x_i\}$ be a future-directed monotonic causal sequence such that every infinite
subsequence $\{y_j\}$ converges to $x$. Then since $\bigcup_j I^-(y_j) \subset \bigcup_{i}
I^-(x_i)$, $I^-(x)\subset \bigcup_{i}I^{-}(x_i)$. On the other hand for every $x_k$, there exists an
$l$ such that $x_k \prec y_l $, so that $I^-(x_k)\subset I^-(y_l)\subset I^-(x)$ which implies that
$\bigcup_{i}I^{-}(x_i)\subset  I^-(x)$, so that $\{x_i\}$ also causally converges to $x$. 
The proof proceeds in an analogous manner for past-directed monotonic causal sequences. \hfill
$\qed$ 

\begin{lemma}\label{ccmc}  
A causal spacetime $(M,g)$ is future and past distinguishing iff  causal convergence
of every monotonic causal sequence $\{x_i \}$ to $x$
implies its manifold converge to $x$, for all $x \in M$.  
\end{lemma} 

\noindent {\bf Proof:} Let $(M,g)$ be past and future distinguishing. Let $\{ x_i\}$ be a future
directed monotonic causal sequence which causally convergences to $x$. If $\{x_i \}$ does not
manifold converge to $x$ then there exists a neighbourhood $O$ of $x$ which contains none of the
$x_i$.  Let $Q$ be a future and past locality neighbourhood of $x$ which lies in $O$ such that $\bar
Q \subset O$. For any $y \in I^-(x,Q)$, there exists an $N$ such that for all $i>N$, $y \in
I^-(x_i)$. (Note that $x_i$ itself does not have to lie in $I^-(x)$.) Since $x_i \nin O$, there
exists $O' \ni x_i$ such that $O' \cap O=\emptyset$, and such that $ O' \in I^+(y)$. If $z \in
I^-(x_i,O')$ then $z \in I^-(x)$ so that $y \pprec z \pprec x$. Thus, there exists a future directed
timelike curve from $y$ to $x$ via a $z \nin Q$ which is a contradiction. Similarly for a past
directed monotononic causal sequence. Thus for a future and past distinguishing spacetime future or
past causal convergence implies manifold convergence. 
       
Conversely assume that every future and past causally convergent sequence also converges in $\mathcal{M}$ to the same
point. Assume wlog that the future distinguishing condition fails at some $x \in M$, so that there exists a $y
\in M$ with $x\neq y$, such that $I^+(x) = I^+(y)$.  Let $\{x_i\}$ be a past-directed monotonic
causal sequence which causally converges to $x$, i.e., $I^+(x) = \bigcup_i I^+(x_i)=I^+(y)$ so that 
$\{x_i\}$ also causally converges to $y$.  Since $\{x_i\}$ causally  
converges to $x$ and $y$, it also converges in $\mathcal{M}$ to $x$ and $y$. But this immediately
leads to a contradiction since $\msM$ is Hausdorff. \hfill $\qed$

As in \cite{fullwood} we can use causal convergence of monotonic sequences to construct a causal
topology $\msP$, namely 
\begin{defn}
A set $O$ is open in $\msP$ if every monotonic causal sequence in $O^c$ which causally converges, also has it's limit point in $O^c$.
\end{defn}

It is easy to establish that $\msP$ is indeed a topology on $M$.

\begin{lemma} 
The collection of sets $\msP$ forms a topology on $M$. 
\end{lemma} 

\noindent \textbf{Proof}: To show that $\msP$ forms a topology on $M$, we need to prove that it
satisfies the three properties of a topology. (i) The proof of $M,\emptyset \in\msP$ is
trivial. (ii) Next let $U_{\alpha}$ be sets in $\msP$, where $\alpha$ belongs to some
indexing set. To show that $\bigcup_{\alpha}U_{\alpha}\in \msP$, note that if a
monotonic causal, causally convergent sequence lies outside $\bigcup_{\alpha}U_{\alpha}$, then it
must lie outside each of the $U_{\alpha}$. Since $U_{\alpha}\in \msP$ for all
$\alpha$, the limit point of the sequence must also lie outside $U_{\alpha}$ for all
$\alpha$. Thus, the limit point is not contained in $\bigcup_{\alpha}U_{\alpha}$, which proves
that $\bigcup_{\alpha}U_{\alpha}\in \msP$. (iii) To show that
$\bigcap_{i=1}^nU_{i}\in \msP$, let $\{x_i\}$ be a monotonic causal, causally
convergent sequence in $M-\bigcap_{i=1}^nU_{i}=\bigcup_{i=1}^n(M-U_{i})$. Since the $(M-U_{i})$'s
are finite in number, at least one of them, say $(M-U_{k})$ should contain a subsequence $\{y_j\}$
of $\{x_i\}$. But since $U_{k}\in \msP$, the limit point of $\{y_j\}$ (which is also
the limit point of $\{x_i\}$ from Lemma \ref{subseq}) must lie in $(M-U_{k})$. Thus, the limit point of
$\{x_i\}$ is in $\bigcup_{i=1}^n(M-U_{i})=M-\bigcap_{i=1}^nU_{i}$, and $\bigcap_{i=1}^nU_{i}\in
\msP$. \hfill $\qed$

\begin{figure}[ht] \label{fatcone} 
\centering \resizebox{1.5in}{!}{\includegraphics{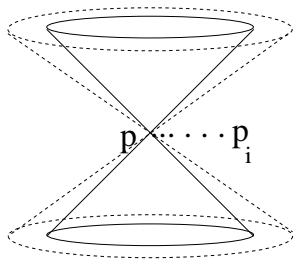}}
\vspace{0.5cm}
\caption{{\small Manifold convergence does not imply causal convergence. Here, we show a double cone
    region emanating from $p$ which is scored by spacelike geodesics. The complement of this
    spacelike cone excludes $p$ and is closed with respect to the $\msP$ topology but as the
    sequence shown above makes clear, it is not closed in the $\msM$ topology.}}\label{figthree}
\end{figure}

\begin{lemma} 
If $(M,g)$ is future \&\ past distinguishing, then $\msM \subset  \msP \subset
\msF$.
\end{lemma} 
\noindent \textbf{Proof}: (a) First we show that $\msM \subset \msP$. Let $O$ be an $\msM$-open
set. Consider a monotonic causal sequence $\{x_i\}$ in $O^c$ which future or past causally converges to $x$. Since
$M$ is future and past distinguishing, $\{x_i\}$ must also converge to $x$ in $\msM$ by Lemma
\ref{ccmc} and hence $x \in O^c$. Since this is true for all future or past causally converging 
monotonic causal sequences in $O^c$, $O$ is also $\msP$-open. Figure \ref{fatcone} shows an example
of a set whose complement is closed in $\msP$, but it is not, in an obvious way, closed in
$\msM$. Thus, $\msM \subset \msP$.    
(b) To show that $\msP \subset \msF$, consider a $V$ which is $\msP$-open. Every monotonic causal sequence, 
and therefore every monotonic timelike sequence in $V^c$ which causally converges to the past or the
future has it's
limit point in $V^c$. Thus, $V$ is $\msF$-open. Conversely, consider the $\msF$-open set $U=<x,p,y>$,
and let $\gamma$ be a past directed null geodesic from $p$. Choose a monotonically increasing causal
$\{ p_i\} $ on $\gamma$ such that that $\{ p_i\}$ casually converges to $p$. Clearly $\{p_i\} \in
U^c$ but $p \in U$ which means that $U$ is not $\msP$-open. \hfill $\qed$ 

\noindent {\bf Acknowledgements:} We would like to thank Fay Dowker for discussions and a careful
reading of an earlier draft. This research was supported in part by the Royal Society International
Joint Project 2006-R2, an NSERC Discovery grant to the McGill University High Energy Theory Group
and an ONR grant to McGill university.

\end{document}